\documentclass[aj]{emulateapj}

\shorttitle{The born-again planetary nebula A78: an X-ray twin of A30}
\shortauthors{Toal\'{a} et al.}

\usepackage{rotating}

\begin{document}

\title{The born-again planetary nebula A78: an X-ray twin of A30}


\author{J.A.\,Toal\'{a}$^{1}$}
\author{M.A.\,Guerrero$^{1}$}
\author{H.\,Todt$^{2}$}
\author{ W.-R.\,Hamann$^{2}$}
\author{Y.-H.\,Chu$^{3,\dagger}$}\thanks{$^{\dagger}$Now at the Institute of Astronomy and Astrophysics, Academia Sinica (ASIAA), Taipei 10617, Taiwan}
\author{R.A.\,Gruendl$^{3}$}
\author{D.\,Sch\"{o}nberner$^{4}$}
\author{L.M.\,Oskinova$^{2}$}
\author{R.A.\,Marquez-Lugo$^{1}$}
\author{X.\,Fang$^{1}$}
\author{G.\,Ramos-Larios$^{5}$}
 
\affil{$^{1}$Instituto de Astrof\'\i sica de Andaluc\'{i}a IAA-CSIC, Glorieta de la Astronom\'\i a s/n, 18008 Granada, Spain; \underline{toala@iaa.es}}
\affil{$^{2}$Institut f\"{u}r Physik und Astronomie, Universit\"{a}t Potsdam, Karl-Liebknecht-Str. 24/25, D-14476 Potsdam, Germany}
\affil{$^{3}$Department of Astronomy, University of Illinois, 1002 West Green Street, Urbana, IL 61801, USA}
\affil{$^{4}$Leibniz-Institut F\"{u}r Astrophysik Potsdam (AIP), An der Sternwarte 16, 14482 Potsdam, Germany}
\affil{$^{5}$Instituto de Astronom\'{i}a y Meteorolog\'{i}a, Av.\,Vallarta No.\,2602, Col. Arcos Vallarta, 44130 Guadalajara, Mexico}

\begin{abstract}
  We present the \textit{XMM-Newton} discovery of X-ray emission from
  the planetary nebula (PN) A78, the second born-again PN detected in
  X-rays apart from A30.  These two PNe share similar spectral and
  morphological characteristics: They harbor diffuse soft X-ray
  emission associated with the interaction between the H-poor ejecta
  and the current fast stellar wind, and a point-like source at the
  position of the central star (CSPN). We present the spectral
  analysis of the CSPN, using for the first time a NLTE code for
  expanding atmospheres which takes line blanketing into account for
  the UV and optical spectra. The wind abundances are used for the
  X-ray spectral analysis of the CSPN and the diffuse emission. The
  X-ray emission from the CSPN in A78 can be modeled by a single
  C\,{\sc vi} emission line, while the X-ray emission from its diffuse
  component is better described by an optically thin plasma emission
  model with temperature $kT$=0.088 keV ($T\approx$1.0$\times$10$^6$
  K).  We estimate X-ray luminosities in the 0.2--2.0~keV energy band
  of $L_{\mathrm{X,CSPN}}$=(1.2$\pm$0.3)$\times$10$^{31}$ erg~s$^{-1}$
  and $L_{\mathrm{X,DIFF}}$=(9.2$\pm$2.3)$\times$10$^{30}$
  erg~s$^{-1}$ for the CSPN and diffuse components, respectively.
\end{abstract}


\maketitle

\section{INTRODUCTION}

Born-again planetary nebulae (PNe) are thought to have experienced a
\textit{very late thermal pulse}\,(VLTP) when the central star (CSPN)
was on the white dwarf (WD) track.  The VLTP event occurs when the
thermonuclear burning of hydrogen in the stellar envelope has built up
a shell of helium with the critical mass to ignite its fusion into
carbon and oxygen
\citep{Herwig1999,Lawlor2006,Millerbertolami2006a,Millerbertolami2006b}. As
the WD envelope is shallow, the increase of pressure from this last
helium shell flash leads to the ejection of newly processed material
inside the old PN, leaving the stellar core intact. As the stellar
envelope expands, its effective temperature decreases and the star
goes back to the asymptotic giant branch (AGB) region in the HR
diagram. The subsequent stellar evolution is fast and will return the
star back to the post-AGB track in the HR diagram
\citep[e.g.][]{Millerbertolami2006b}: the envelope of the star
contracts, its effective temperature and ionizing photon flux
increase, and a new fast stellar wind develops.  This canonical model,
however, has notable difficulties to reproduce the relatively low C/O
ratio and high neon abundances found in born-again PNe
\citep[e.g.,][]{Wesson2003,Wesson2008}. Alternative scenarios,
invoking the possible evolution through a binary system or a nova
event immediately after the late helium shell flash, have been
discussed by \citet{Lau2011}, although none of them are completely
satisfactory.

The born-again phenomenon is rare, with A30, A58 (Nova Aql 1919), A78,
and the Sakurai's object (V\,4334 Sgr) being the most studied objects
of this class. These PNe harbor complex physical processes: the
hydrogen-poor material ejected by the star during the born-again event
will be photoevaporated by the ionizing photon flux from the CSPN and
swept up by the current fast stellar wind
\citep[see][]{Guerrero2012,Fang2014}. These objects evolve very fast
after the VLTP, thus, they provide a rare opportunity to study such
complex phenomena and their real-time evolution
\citep[e.g.,][]{Evans2006,Hinkle2008,Clayton2013,Hinkle2014}.

\begin{figure*}
\begin{center}
\includegraphics[angle=0,width=0.33\linewidth]{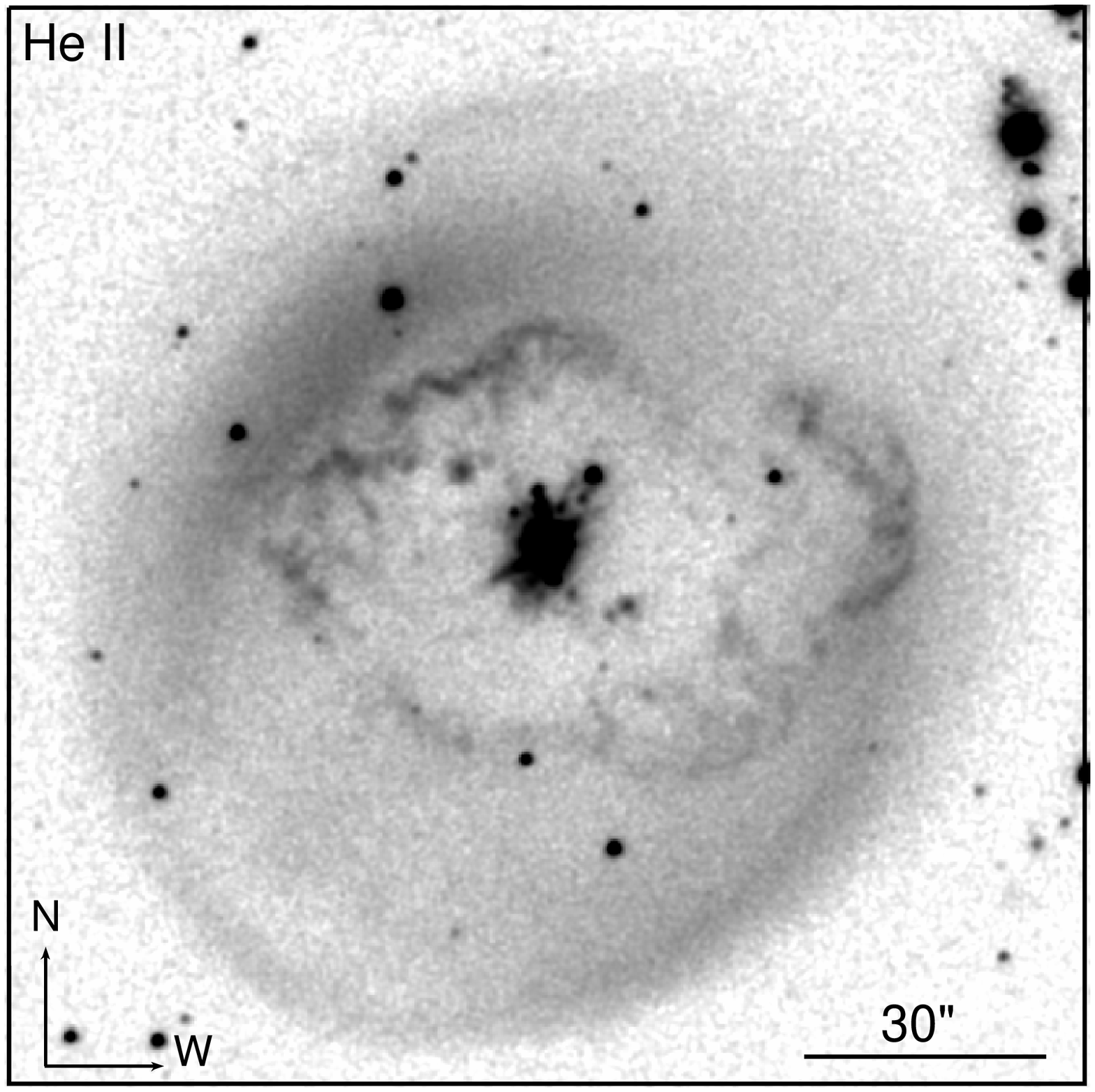}~
\includegraphics[angle=0,width=0.33\linewidth]{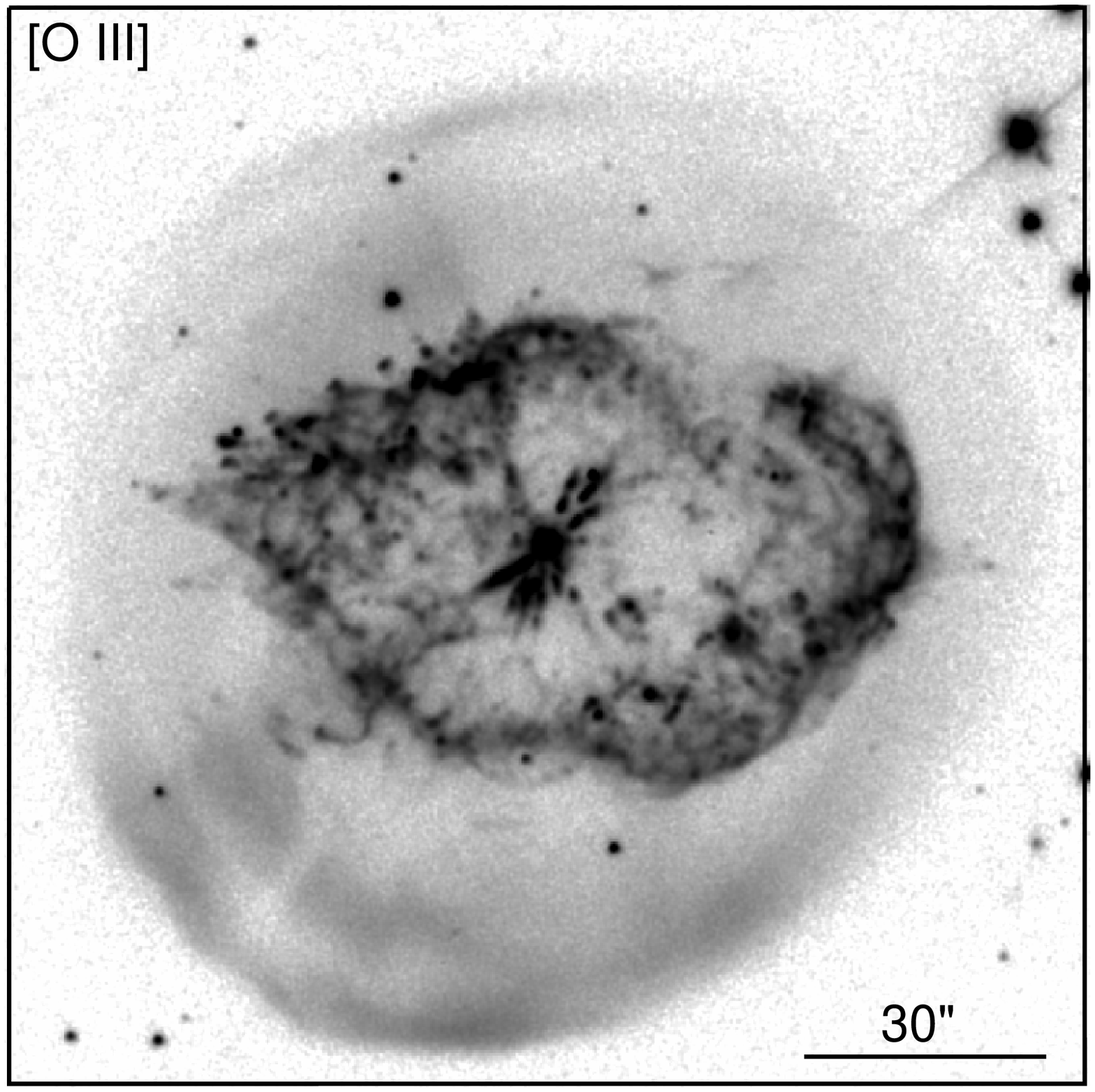}~
\includegraphics[angle=0,width=0.33\linewidth]{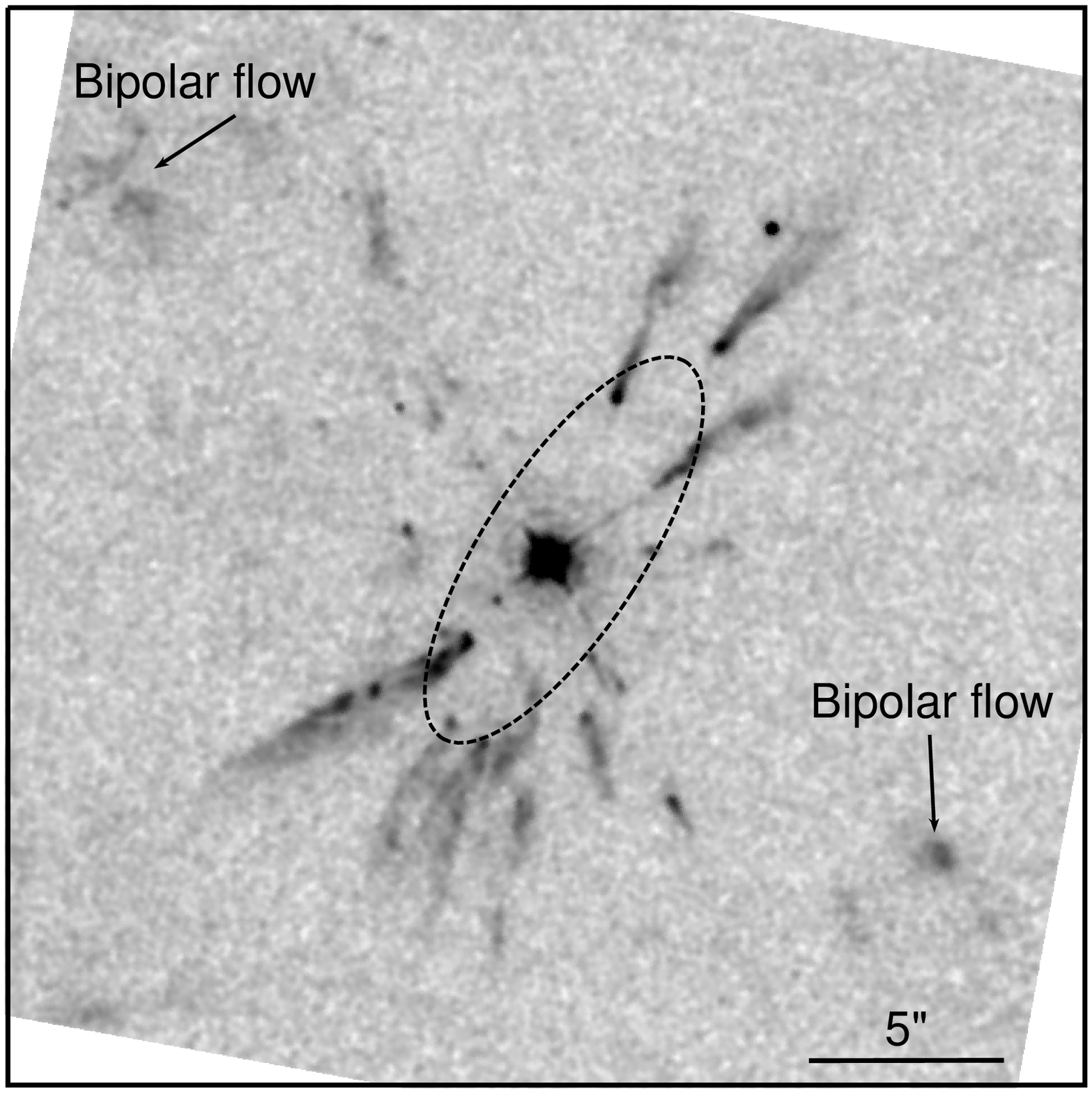}
\caption{He\,{\sc ii} (left) and [O\,{\sc iii}] (center) narrow-band
  images of A78 {\bf as obtained at the Nordic Optical Telescope (see
  Section~3 for details),} and (right) \emph{HST} WFPC2 F502N image of the central
  part of A78. The equatorial plane and the bipolar flows are marked
  with a dashed-line ellipse and arrows, respectively.}
\end{center}
\label{fig:a78_optic}
\end{figure*}

Among these PNe, A30 and A78 share similar characteristics suggesting
that they are at a comparably late stage after the born-again
event. They exhibit large ($\sim$2\arcmin\ in size) limb-brightened
outer shells that surround an ensemble of H-poor clumps that are
prominent in [O~{\sc iii}] narrow band images \citep[][see
Figure~1]{Jacoby1979}. The outer hydrogen-rich shells are ellipsoidal
in shape and expand at $\sim$40~km~s$^{-1}$, while the H-deficient
knots detected in [O~{\sc iii}] have velocities up to 200~km~s$^{-1}$
\citep{Meaburn1996,Meaburn1998}. The central parts in A30 and A78
were imaged by the \emph{Hubble Space Telescope}
\citep[\textit{HST};][]{Borkowski1993,Fang2014}. The detailed
\emph{HST} WFPC2 [O~{\sc iii}] images revealed cometary structures
distributed on an equatorial plane (a few arcsec from the star) and
polar features, which, in the case of A78, are more diffuse and are
located at $\sim$12--14\arcsec\ from its CSPN (see
Figure~\ref{fig:a78_optic}-right panel).

X-ray emission has been detected within a number of PNe, including A30
\citep[e.g.,][and references
therein]{Guerrero2012,Kastner2012,Ruiz2013,Freeman2014}.  This
born-again PN has been studied with \textit{ROSAT} (PSPC and HRI),
\textit{Chandra}, and \textit{XMM-Newton} X-ray satellites
\citep{Chu1995,Chu1997,Guerrero2012}.  Its X-ray emission originates
from the CSPN, but there is also diffuse emission spatially coincident
with the cloverleaf-shaped H-poor structure detected in [O~{\sc iii}].
The X-ray emission from both the CSPN and the diffuse extended
emission are extremely soft.

The X-ray properties of A30 and the similarities of this nebula to A78
motivated us to obtain \emph{XMM-Newton} observations, as the only
previous X-ray observations of A78 by \emph{Einstein} yielded a rather
insensitive upper limit \citep{Tarafdar1988}. In this paper we present
the analysis of new \emph{XMM-Newton} observations that reveal the
existence of hot gas associated with the H-poor knots inside the
eye-shaped inner shell of A78, and also a point-like source of X-ray
emission at its CSPN. The outline of this paper is as follows.  The
stellar wind properties and abundances of A78 CSPN are derived in \S2.
The \emph{XMM-Newton} observations are presented in \S3, and the
spatial distribution and spectral properties of the X-ray emission in
\S4 and \S5, respectively.  A discussion is presented in \S6 and we
summarize our findings in \S7.

\section{NLTE analysis of the central star}
 
We analyzed the optical and UV spectra of the CSPN of A78 using the
most recent version of the Potsdam Wolf-Rayet (PoWR) model
atmosphere\footnote{\url{http://www.astro.physik.uni-potsdam.de/PoWR}}.
The PoWR solves the NLTE radiative transfer problem in a spherical
expanding atmosphere simultaneously with the statistical equilibrium
equations and at the same time accounts for energy
conservation. Iron-group line blanketing is treated by means of the
superlevel approach \citep{Grafener2002}, and a wind clumping in
first-order approximation is taken into account \citep{Hamann2004}. We
did not calculate hydrodynamically consistent models, but assumed a
velocity field following a $\beta$-law with $\beta = 1$. We also
performed tests with different $\beta$-laws, e.g., $\beta=0.8$, $\beta
=2$, and a double-$\beta$ law, but we found the impact of different
$\beta$-laws to be much smaller than the change of other parameters,
such as effective temperature and mass-loss rate. Our computations
applied here include complex atomic models for helium, carbon,
nitrogen, oxygen, neon, fluorine, hydrogen, and the iron-group
elements.

The synthetic spectrum was corrected for interstellar extinction due
to dust by the reddening law of \cite{Cardelli1989}, as well as for
interstellar line absorption for the Lyman series in the UV range.

\begin{deluxetable*}{lcl}[ht]
\tablewidth{0.5\textwidth}
\tablecaption{Parameters of the CSPN of A78}
\tablehead{
\multicolumn{1}{l}{Parameter} &
\multicolumn{1}{c}{Value}  &
\multicolumn{1}{l}{Comment} 
 }
\startdata
$T_{\mathrm{eff}}$\,(kK)\tablenotemark{a}         &  117   &  \\
log($L/L_{\odot}$)              &  3.78   & Adopted \\
$M_{*}$\,($M_{\odot}$)           &  0.6   & Adopted \\
$R_{*}$\,($R_{\odot}$)\tablenotemark{b}          & 0.19    & $R_{*}\propto L^{1/2}$ \\
$R_{\tau=2/3}$\,($R_{\odot}$)                     & 0.19    &  \\
$v_{\infty}$\,(km~s$^{-1}$)      &  3100   &  \\
$D$ (clumping factor)          & 10      & Adopted \\
log$\dot{M}$~($M_{\odot}$~yr$^{-1}$) & -7.8 & $\dot{M}\propto D^{-1/2}L^{3/4}$\\
$d$\,(kpc)                     &  1.40   &  $d\propto L^{1/2}$\\
$E_{B-V}$\,(mag)                & 0.12    & \\
\hline
  & Abundances (mass fraction) &  \\
\hline
He & 0.55 &  \\
C  & 0.30 &  \\
N  & 0.015 & \\
O  & 0.10  & \\
Ne & 0.04 & \\
F  & $1.3 \times 10^{-5}$ & \\ 
Fe-group & 1.4$\times$10$^{-4}$ & 
\enddata
\tablenotetext{a}{$T_\mathrm{eff}$ is defined as the effective
  temperature at the radius $R_{*}$.}  
\tablenotetext{b}{The stellar
  radius $R_{*}$ refers to, by definition, the point where the radial
  Rosseland optical depth is 20.}
\label{tab:CSPN} 
\end{deluxetable*}

\begin{figure*}
\includegraphics[width=\textwidth]{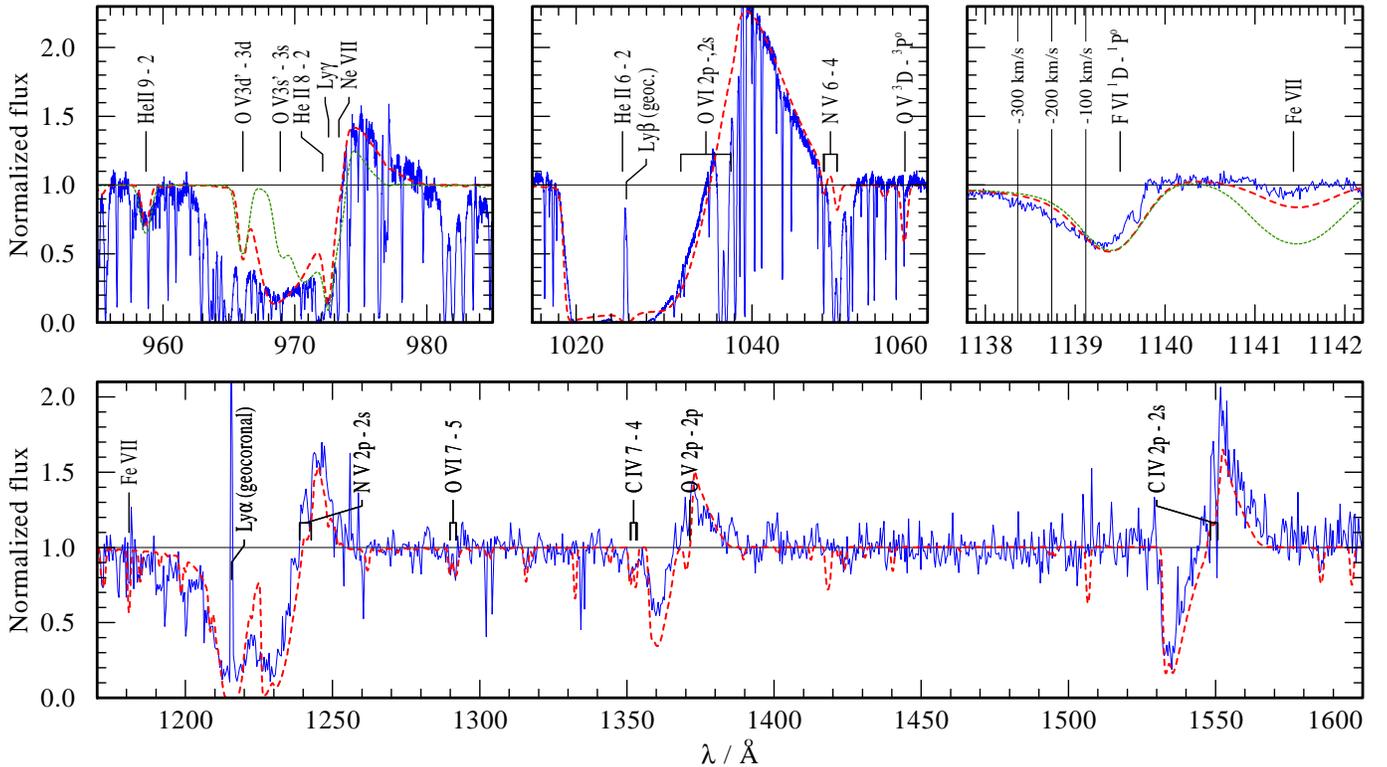}
\caption{Details of the normalized UV spectrum of A78. Observations
  (blue thin line) from {\em FUSE} (upper panels) and {\em IUE} (lower
  panel) are shown together with the synthetic spectrum of our model
  (dashed lines) with parameters as in Table~\ref{tab:CSPN}. The model
  spectrum was convolved with a Gaussian with $0.5$~\AA\, FWHM to
  match the resolution of the observation.  {\em Top-left panel:} A
  model with 4\% Ne (red dashed) compared to a model with only solar
  Ne abundance (green dashed).  {\em Top-middle panel}: Best-fit model
  with $v_\infty=3100$~km~s$^{-1}$ plus depth dependent
  microturbulence (red dashed).  {\em Top-right panel}: Best-fit model
  to the Fe~{\sc vii} 1141.43 \AA\ line with 10\% solar Fe abundance
  (red dashed), compared to a model with solar Fe abundance (green
  dashed).  This panel also includes the F~{\sc vi} 1139.5 \AA\ line,
  where the vertical lines in the \emph{FUSE} spectrum indicate
  different Doppler shifts with respect to the F~{\sc vi} line. {\em
    Bottom panel}: Our best-fit model (red dashed) vs. {\em IUE}
  observation.}
\label{fig:uv}
\end{figure*}

\begin{figure*}
\includegraphics[width=\textwidth]{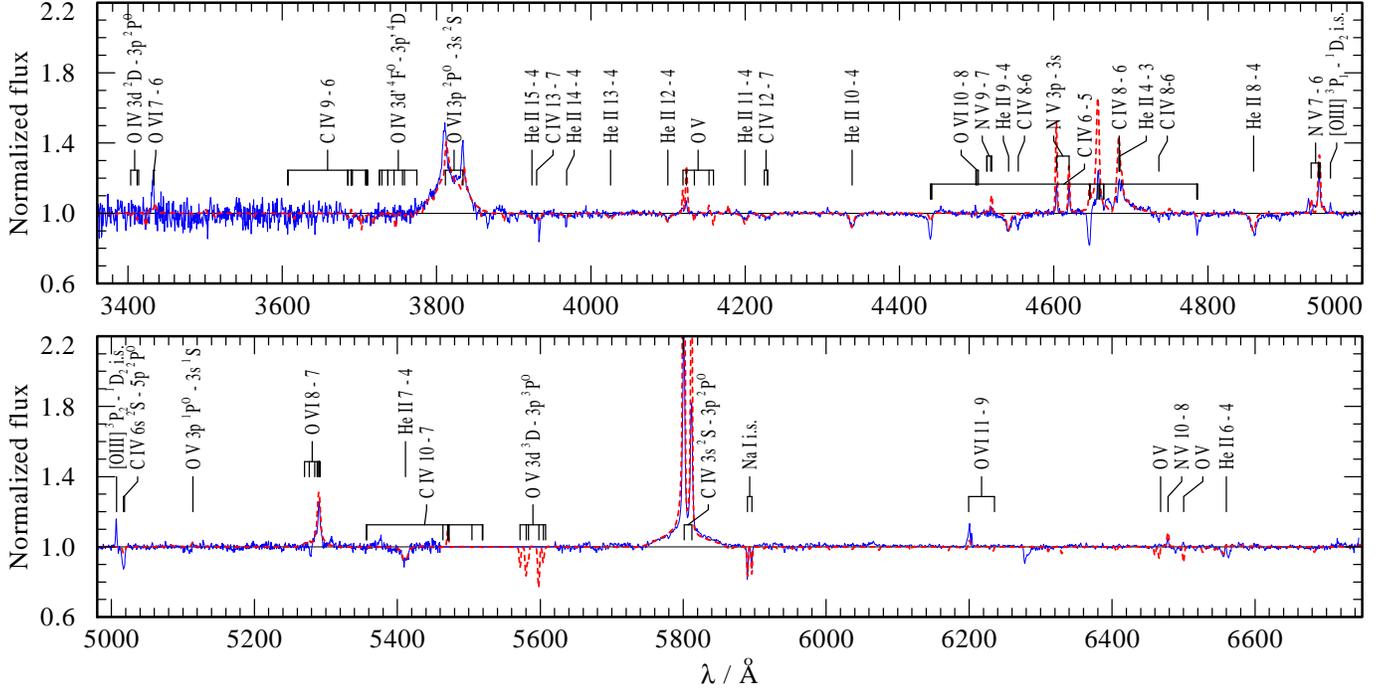}
\caption{Optical spectrum of the central star of A78. Normalized
  observation (blue) vs.\ synthetic spectrum of our best-fit model
  (red dashed). The model spectrum was convolved with a Gaussian with
  $2\,$\AA\ FWHM to match the resolution of the observation, inferred
  from the interstellar Na\,{\sc i} doublet.}
\label{fig:opt}
\end{figure*}

\begin{figure*}
\begin{center}
\includegraphics[width=0.5\textwidth]{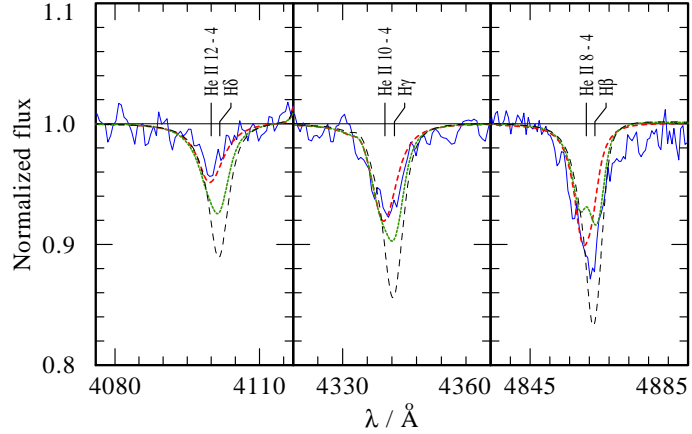}
\caption{Details of the optical spectrum of A78: Observation (blue
  thin solid lines) vs.\ models without hydrogen (red thick dashed),
  with 10\% H (green thick dotted), and with 30\% H (black thin
  dashed).}
\end{center}
\label{fig:hydrogen}
\end{figure*}

\begin{figure*}
\includegraphics[width=\textwidth]{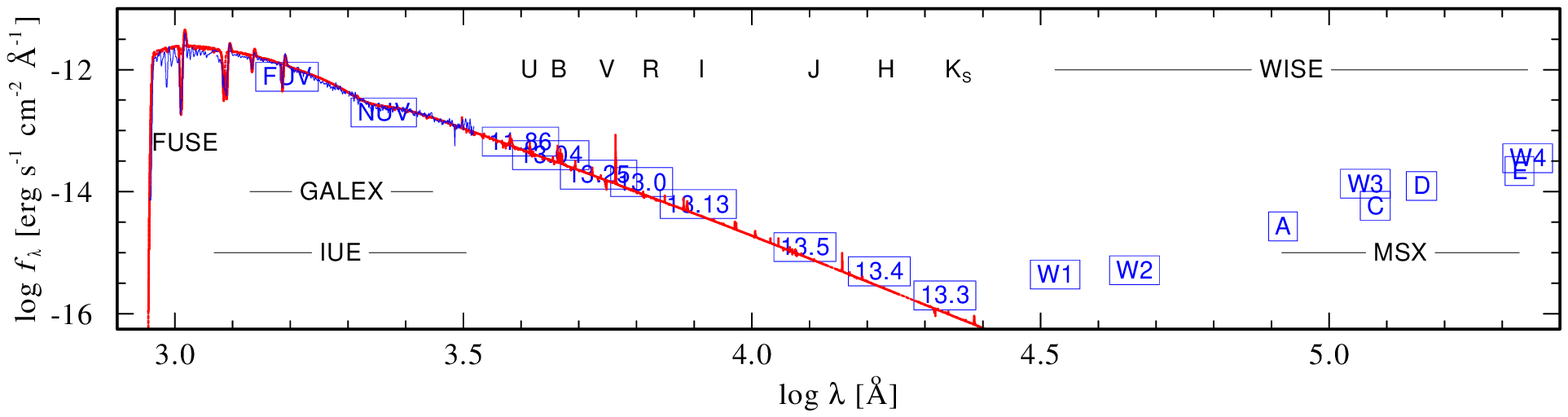}
\caption{Spectral energy distribution (SED) of the central star of
  A78 from the UV to the infrared range. Observations (blue) are
  photometric measurements in the indicated bands and calibrated {\em IUE} 
  and {\em FUSE} UV spectra. 
  The theoretical SED derived from our stellar model with the parameters
  compiled in Table~\ref{tab:CSPN} is also shown.}
\label{fig:sed}
\end{figure*}

The UV spectrum of the CSPN of A78 has been observed by the {\em Far
  Ultraviolet Spectroscopic Explorer (FUSE)} and the {\em
  International Ultraviolet Explorer (IUE)} satellites.  Data from
these observations have been retrieved from MAST, the Mikulski Archive
for Space Telescopes at the Space Telescope Science
Institute\footnote{STScI is operated by the Association of
  Universities for Research in Astronomy, Inc., under NASA contract
  NAS5-26555.}.  We used the {\em FUSE} observation with ID
e1180101000 (PI: J.\ Kruk) obtained on 2004 November 11 in the
spectral range 916-1190~\AA\, for a total exposure time of 58\,ks. The
{\em IUE} observations in the spectral range 1150-1975\,\AA\ consists
of two datasets with IDs SWP19879 and SWP19906, both taken with the
large aperture at high dispersion with total exposure times of
25.5\,ks and 25.2\,ks, respectively.  Only low dispersion {\em IUE}
observations were available for the spectral range 2000-3300\,\AA.
The data set with ID LWP23314LL obtained through the large aperture
for a total exposure time of 600\,s was used.  Optical spectra of 
$R\approx 6000$ were obtained by K.\ Werner using the TWIN spectrograph
on the 3.5m-telescope at Calar Alto in October 1990 and in September
1991 \citep[see][]{Werner1992}.

Since the stellar spectra of the CSPN of A30 \citep{Guerrero2012} and
A78 are similar, the spectral fits resulted in similar wind properties
and abundances (see Table~\ref{tab:CSPN}). For the determination of
the stellar temperature we used mainly the ratio of the line strengths
of O\,{\sc vi} vs.\ O\,{\sc v} wind lines.  The best fit to these
lines was obtained with an effective temperature of $T_*=117 \pm
5\,$kK (see Figures~\ref{fig:uv} and \ref{fig:opt}).

From the strengths of the emission and P Cygni lines we derived the
mass-loss rate under the assumption of a typical central star
luminosity and mass of $L=6000\,L_\odot$ and $M=0.6\,M_\odot$
\citep[see e.g.][]{Schonberner2005,MillerBertolami2007}.  The inferred
value of $\dot{M} = 1.6\times 10^{-8}\,M_\odot$~yr$^{-1}$ is about
half of the value from \cite{Werner1992} and \cite{Leue1993}, also
because we account for clumping and use a different luminosity (see
Table~\ref{tab:CSPN} for scaling relations). Indeed, the mass-loss
rate determined by \cite{Koesterke1998} is the same as ours, if
rescaled to the values of distance and stellar luminosity used in our
analysis.

The blue edge of the P Cygni profiles was used to estimate a terminal
wind velocity of about $3100\pm100$km/s, consistent with that found by
\citet{Guerrero2013}. Additional broadening due to depth-dependent
microturbulence with $v_\mathrm{D}=50$~km~s$^{-1}$ in the photosphere
up to $v_\mathrm{D}=300$~km~s$^{-1}$ in the outer wind was taken into
account and allows to fit the widths of the O\,{\sc vi} and the
C\,{\sc iv} resonance lines simultaneously (see Figure~\ref{fig:uv}).

The strong Ne\,{\sc vii} line at 973.33\,\AA\ \citep{Herald2005} 
observed in the UV spectrum (see Figure~\ref{fig:uv}) can only be 
reproduced by models with a supersolar Ne abundance.
Similarly, the strength of the N\,{\sc v} lines (Figures~2 and 3) 
implies a supersolar nitrogen abundance of 1.5\% by mass.
To reproduce the observed strength of the F\,{\sc vi}\ 1139.5\,\AA\ line 
a fluorine abundance of at least 25$\times$ the solar value is needed, 
similar to what was found by \cite{Werner2005} for the same object as 
well as for other H-deficient post-AGB stars. 
They also mentioned an asymmetry of this line meaning that the line 
is partly formed in the wind, as reproduced by our wind models (see 
Figure~\ref{fig:uv}).

Initial solar abundances of the iron group resulted in Fe~{\sc vii}
lines more intense than those observed in the \emph{FUSE} spectrum
(see Figure~2-top right panel).  Accordingly, the abundance of the
iron group elements had to be reduced down to 1/10 of the solar value
to obtain a consistent fit of the Fe~{\sc vii} lines.  Our iron
estimate is in contrast to that found by \citet{Werner2011}, who
suggested that the strong Fe~{\sc viii} 1148.22 \AA\ line would imply
solar Fe abundances.  They recognize, however, that the Fe~{\sc viii}
line profile, much broader than the prediction from their static NLTE
models, called for an analysis using expanding atmosphere models as
that performed here.  Following the interpretation of the subsolar
Fe/Ni ratio reported in Sakurai's object \citep{Asplund1999}, the iron
deficiency in A78 can be explained by the conversion of iron into
heavier elements by s-process neutron captures.

We also tried to constrain the hydrogen abundance. The best fit to the
Balmer lines is obtained by models without hydrogen. However, at the
given resolution and S/N of our optical observation, a hydrogen
abundance below 10\% by mass would escape detection (Figure~4).

The absolute flux of the model is diluted by the distance to the
central star, which we consider to be a free parameter. We obtain a
consistent fit of the spectral energy distribution from the far UV to
the near infrared range (see Figure~\ref{fig:sed}) for a reddening of
$E_{B-V}=0.12$~mag and a distance of $d=1.4$~kpc. These values are
consistent with the reddening of $E_{B-V}=0.15$~mag and the distance
of $d=1.5\,$kpc reported by \citet{Jeffery1995} and
\citet{Harrington1995}, respectively. We note, however, that our
distance estimate is 30\% smaller than that of \citet{Frew2008}.


\section{Observations}

\textit{XMM-Newton} observed A 78 on 2013 June 3
(Observation ID 0721150101, PI: M.A.\,Guerrero) using the European
Photon Imaging Cameras (EPIC) and Reflective Grating Spectrographs
(RGS). The observations were performed in the Full Frame Mode with the
thin optical filter for a total exposure time of 59.4~ks. The
data were reprocessed with the \textit{XMM-Newton} Science
Analysis Software (SAS) 13.5 with the most up-to-date
\textit{XMM-Newton} calibration files available on the Current
Calibration File as of 2014 January 7. The net exposure times were
59.4, 59.1, 59.1, 59.4, and 59.4~ks for the EPIC-pn, EPIC-MOS1, and
EPIC-MOS2, RGS1, and RGS2, respectively. After processing the
effective times were reduced to 23.9, 43.6, 42.5, 59.1, and 59.0~ks
for the EPIC-pn, MOS1, MOS2, RGS1, and RGS2 respectively.

To help study the distribution of the X-ray emission, we obtained
He~{\sc ii} and [O~{\sc iii}] narrow-band images of A78 on 2014 July
19 using the Andalusian Faint Object Spectrograph and Camera (ALFOSC)
at the Nordic Optical Telescope (NOT).  The central wavelengths and
bandpasses (FWHM) of the filters are 4687 \AA\ and 43 \AA\ for He~{\sc
  ii}, and 5010 \AA\ and 35 \AA\ for [O~{\sc iii}], respectively.  The
images have total exposure times of 1800~s each.  The average seeing
during the observation was $\sim$0\farcs7.  The final processed images
are shown in the left and middle panels of Figure~1.

\section{Spatial distribution of the X-ray emission}

\begin{figure*}
\begin{center}
\includegraphics[angle=0,width=1.0\linewidth]{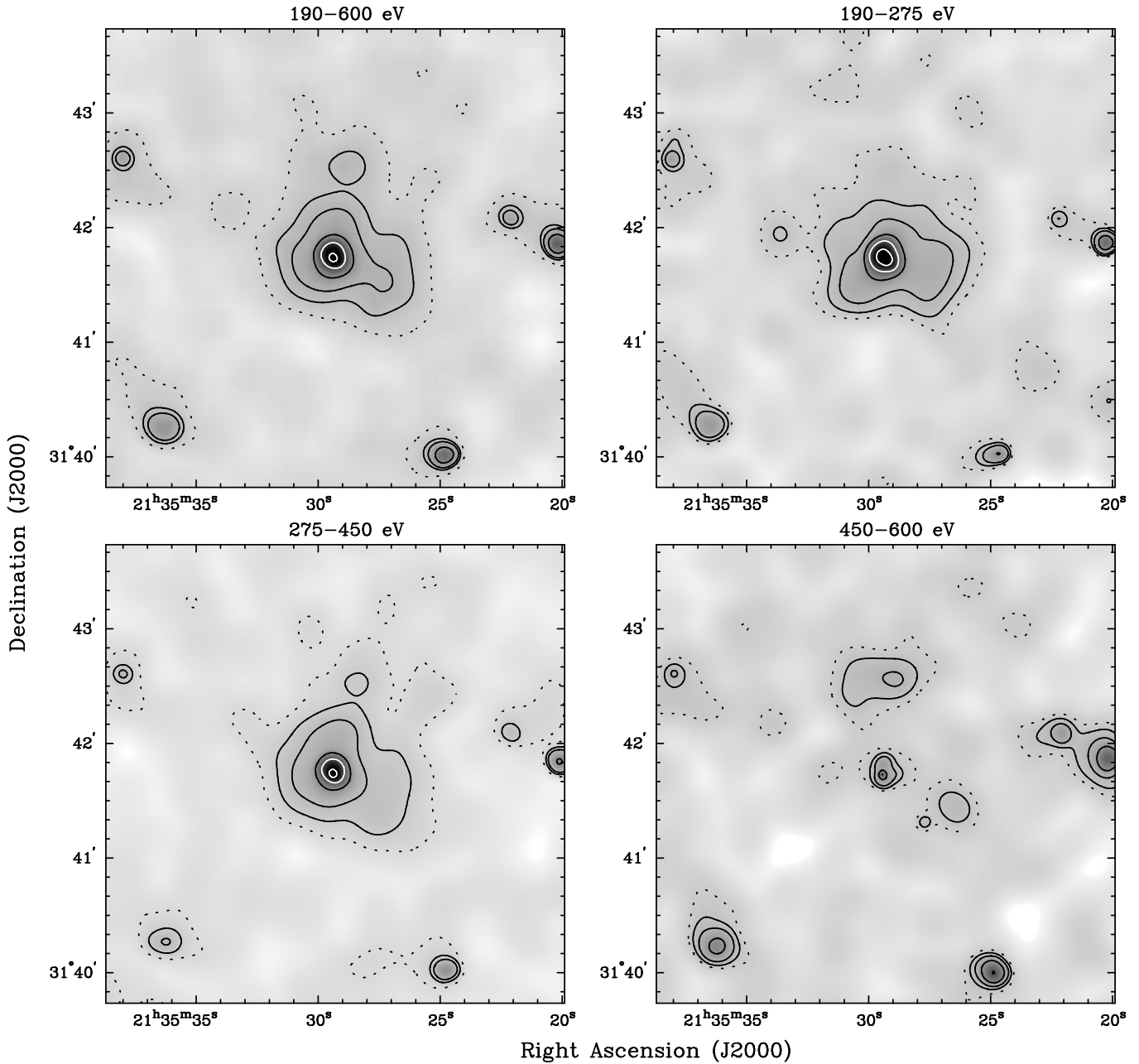}
\caption{Exposure-corrected \textit{XMM-Newton} EPIC images of A78 in
  different bands. The images have a pixel size 1$\arcsec$. The images
  are centered at the central star in A78. Other point-like sources
  are presented on the images. The black lower contours correspond to
  1, 3, 5, and 10$\sigma$ over the background level, while the white
  upper contours represent 20\% and 60\% of the peak intensity.}
\end{center}
\label{fig:a78_cont2}
\end{figure*}

For direct comparison with the \textit{XMM-Newton} observations of
A30, we created EPIC images of A78 in four different energy bands: 
soft 190-275~eV, medium 275--450~eV, hard 450--600~eV, and total
190--600~eV.  Individual EPIC-pn, EPIC-MOS1, and EPIC-MOS2 images were
extracted, merged together, and corrected for exposure maps.  The
final smoothed exposure-corrected images of the four energy bands are
show in Figure~6.

Figure~6 reveals a bright source associated with the CSPN and diffuse
X-ray emission within A78.  Both the point-like source and diffuse
emission seem to fade away at energies $>$450~eV. Other point-like
sources are present in the panels shown in Figure~6, in particular the
point-like X-ray source detected in all energy bands $\sim$37\arcsec\
north of the CSPN of A78. The optical counterpart of this X-ray source
is detected faintly in the optical images in Figure~1 with coordinates
(R.A,Dec.)=(21$^\mathrm{h}$\,35$^\mathrm{m}$\,28.76$^\mathrm{s}$,
$+$31$^{\circ}$\,42$\arcmin$\,31.2$\arcsec$). No counterpart is
identified in the NED and SIMBAD databases. This source is most likely
a background source, as it does not have any morphological correlation
with A78.

In Figure~7 
we compare the spatial extent of the X-ray emission with the optical
H$\alpha$ and [O~{\sc iii}] images presented in Figure~1. The diffuse
X-ray emission in A78 does not fill the elliptical outer shell, but it
seems to be bounded by the [O~{\sc iii}] bright eye-shaped shell, as
it was also the case for the distribution of diffuse X-rays in
A30. Furthermore, there is a local peak in the diffuse X-ray emission
that seems to be associated with a H-poor clump toward the SW
direction from the CSPN. In a similar manner, A30 also presents a
maxima in the X-ray emission associated with the H-poor clumps,
suggesting that the two born-again PNe may have similar origins of
X-ray emission.

\begin{figure}
\begin{center}
\includegraphics[angle=0,width=1\linewidth]{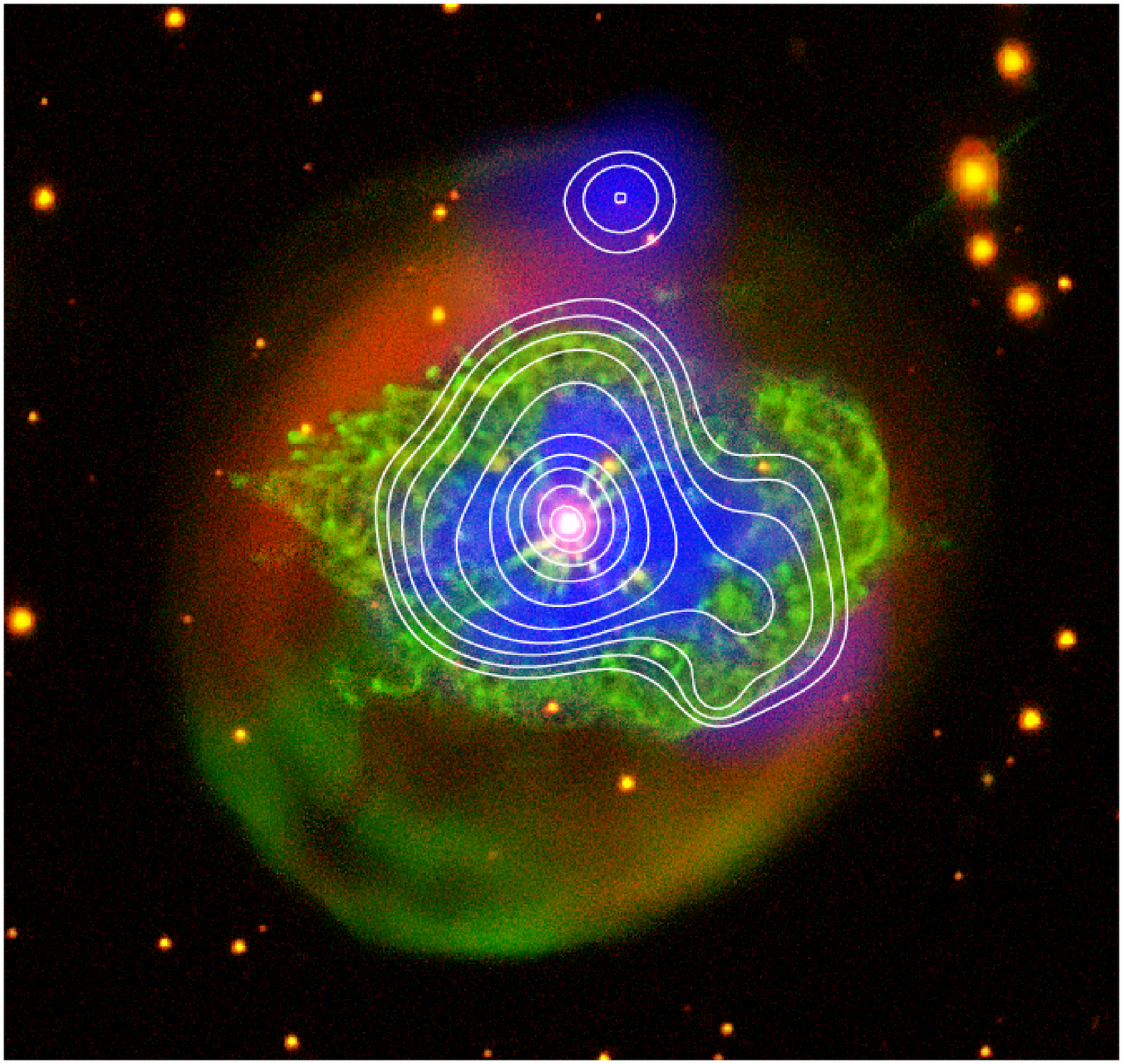}
\caption{Composite color picture of the {\it XMM-Newton} EPIC
  190-600~eV (blue) and NOT ALFOSC [O\,{\sc iii}] (green) and He\,{\sc
    ii} (red) images of A78. To emphasize the comparison between the
  spatial distributions of the X-ray-emitting gas and nebular
  component in A78, X-ray contours of the same energy band have been
  overplotted. The emission to the north of A78 corresponds to a
  point-like X-ray source in the field of view of the observations
  (see text).}
\end{center}
\label{fig:A78_oiii_xray}
\end{figure}

\begin{figure}
\begin{center}
\includegraphics[angle=0,width=1\linewidth]{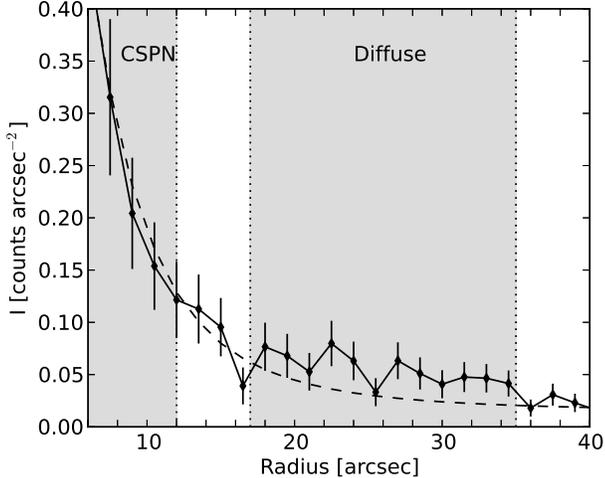}
\caption{EPIC-pn radial profile of the X-ray emission from A78
  extracted using \textit{eradial}. The dashed line is the fitted PSF
  to the radial profile.}
\end{center}
\label{fig:A78_eradial}
\end{figure}

To better assess the extent and intensity of the diffuse X-ray
emission, we have used the SAS task \textit{eradial} to extract a
radial profile of the X-ray emission centered on the CSPN of A78 and
compare it to the theoretical point-spread function (PSF) of the
observation.  This comparison is shown in Figure~8, where the PSF
scaled to the radial profile fits nicely the emission from the CSPN
until a radius of 12\arcsec. An excess of diffuse emission is
detected above the PSF profile between 17--35\arcsec.
To estimate the contribution of the CSPN emission to the diffuse
emission component, we have integrated the radial profile emission
obtained with \textit{eradial} for distances smaller than 34\arcsec,
and computed the percentage to the total emission from the CSPN using
the PSF model.  The contribution of the CSPN to the total emission is
99\% for radial distances $<$12\arcsec, 79\% for distances
$<$34\arcsec, and only 24\% for distances between 17\arcsec\ and
34\arcsec.

\section{Spectral properties of the X-ray emission}

To study the global spectral properties of the X-ray emission from
A78, we have extracted the EPIC-pn, EPIC-MOS, and combined RGS1+RGS2
spectra shown in Figure~9. 
The EPIC spectra
have been extracted from an elliptical region that encloses the whole
emission from A78.  These spectra (Figure~9-top) are very soft and
resemble those presented by \citet{Guerrero2012} for A30. The EPIC-pn
spectrum peaks at 0.3-0.4~keV with a rapid decay at energies greater
than 0.5~keV. This spectrum shows evidence for an emission line at
$\sim$0.58~keV, absent in the EPIC-pn spectrum of A30, which would 
correspond to the O\,{\sc vii} triplet.  The count rates for
the EPIC-pn, EPIC-MOS1, and EPIC-MOS2 in the 0.2--2.0~keV energy range
are 18.2, 2.4, and 2.1 counts~ks$^{-1}$, respectively.

\begin{figure}
\begin{center}
\includegraphics[angle=0,width=1.05\linewidth]{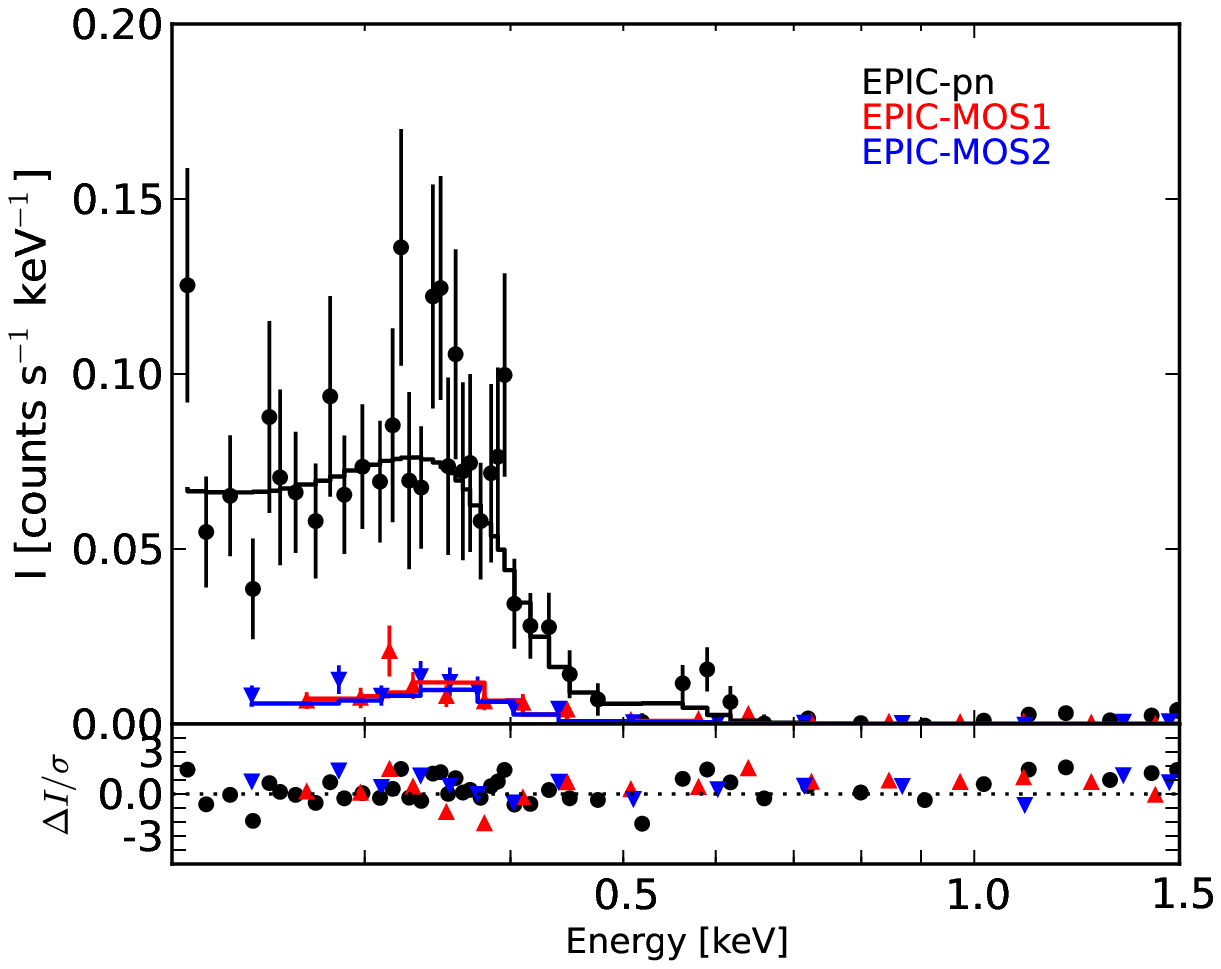}\\
\includegraphics[angle=0,width=1.05\linewidth]{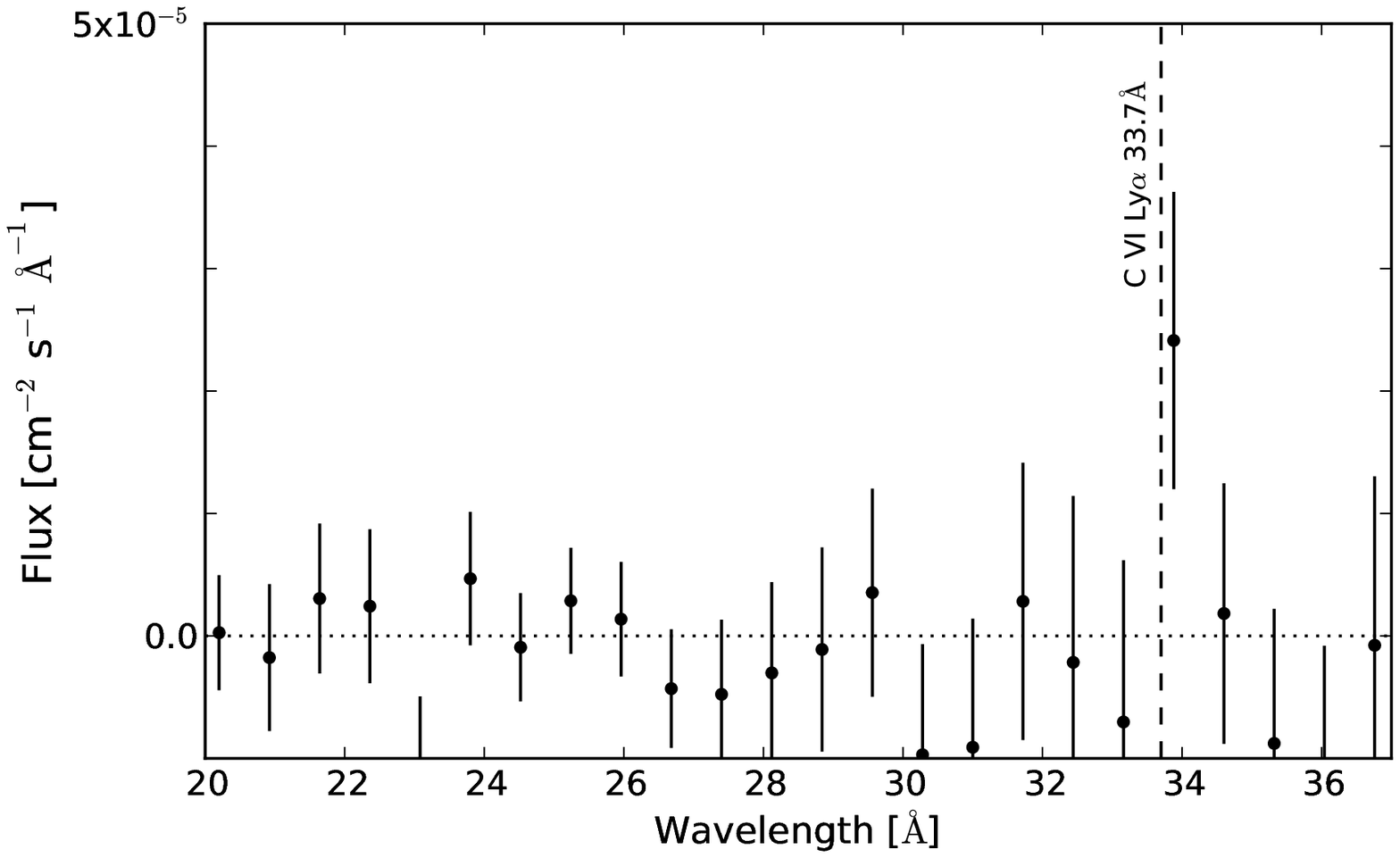}
\caption{\emph{XMM-Newton} EPIC (top) and combined RGS1+RGS2 (bottom)
  background-subtracted spectra of A78. The best-fit models to the
  EPIC-pn and MOS cameras are shown with solid lines in the top panel
  (see \S\,5.1). Residuals of these fits are also shown. The RGS
  spectrum displays the C\,{\sc vi} emission line at
  33.7~\AA(=0.37~keV).}
\end{center}
\label{fig:spec_todo}
\end{figure}

The combined RGS1+RGS2 background-subtracted spectrum of A78 can help
us identify the emission detected around 0.3-0.4~keV in the EPIC-pn
camera.  Figure\,9-bottom panel shows that this is mostly due to the
C~{\sc vi} emission line at 33.7~\AA\ (=0.37~keV).  However, because
of the low MOS and RGS count rates, we will mainly focus on the
spectral analysis from the spectrum extracted from the EPIC-pn camera
for further discussion.

We have extracted separately EPIC-pn spectra for the CSPN and for the
diffuse X-ray emission. The spectrum from the CSPN was extracted using
a circular aperture of radius 12\arcsec\, centered at the position of
the star with a background extracted from regions with no contribution
of diffuse emission. The spectrum from the diffuse emission was
extracted using an elliptical aperture that covered the extension of
the [O~{\sc iii}] filamentary shell with a minor axis of 34\arcsec\ to
avoid contamination from the point-like source to the North.
According to \S4, a circular region of radius 17\arcsec\ centered at
the position of the CSPN was excised to reduce the contamination from
the CSPN. The resultant background-subtracted EPIC-pn spectra of the
CSPN and that of the diffuse component, shown in Figure~10, have net
count rates of 8.8 and 8.0~counts~ks$^{-1}$, respectively.

Figure~10 shows subtle differences in the spectral shapes from the CSPN
and diffuse emission. For example, the spectrum from the diffuse
component (Figure~10-right) shows the spectral line at $\sim$0.58~keV,
whilst that of the CSPN (Figure~10-left) does not. As for A30, both
spectra seem to peak at the energy $\sim$0.37~keV of the C~{\sc vi}
line.

\begin{figure*}
\begin{center}
\includegraphics[angle=0,width=0.5\linewidth]{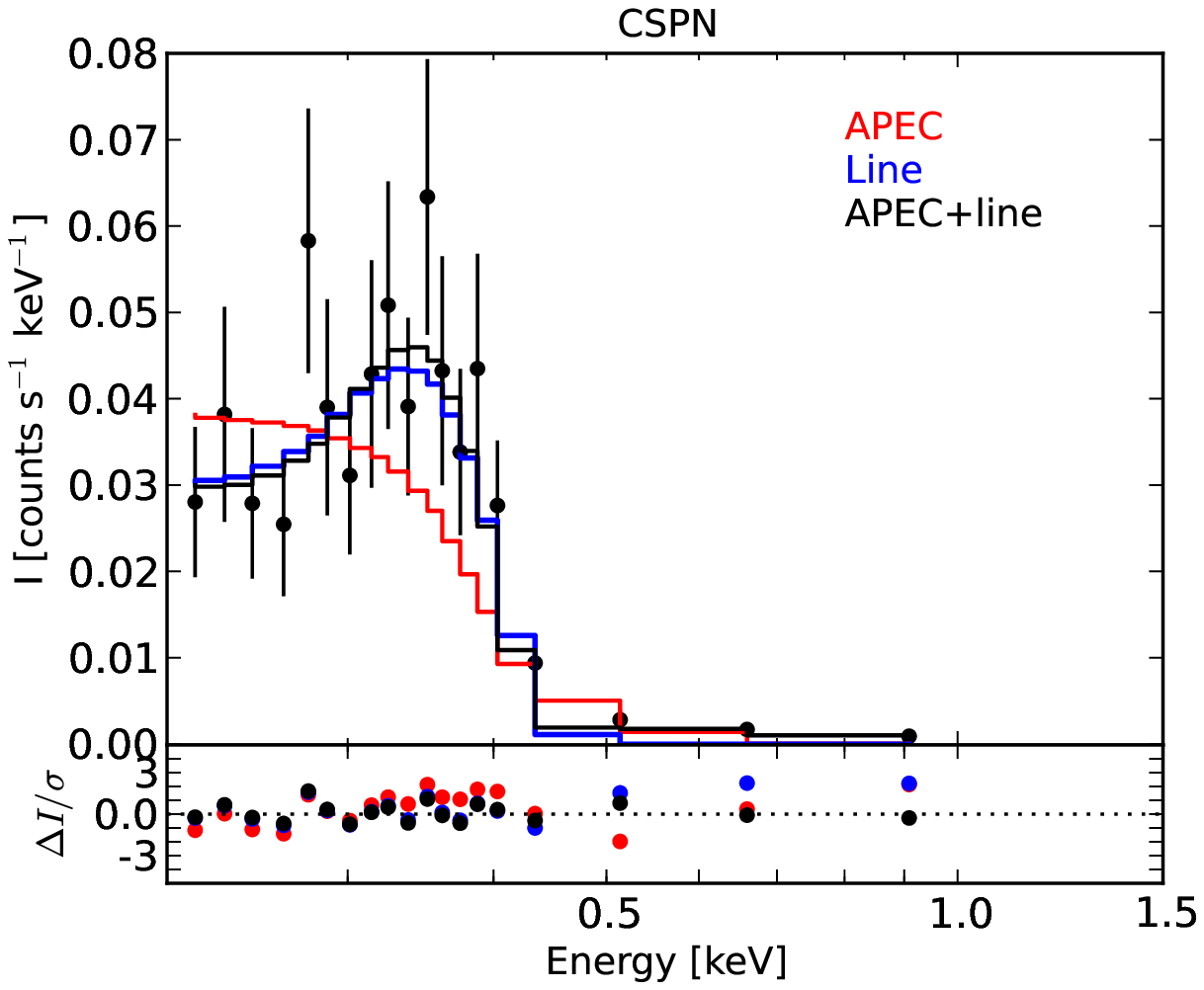}~
\includegraphics[angle=0,width=0.5\linewidth]{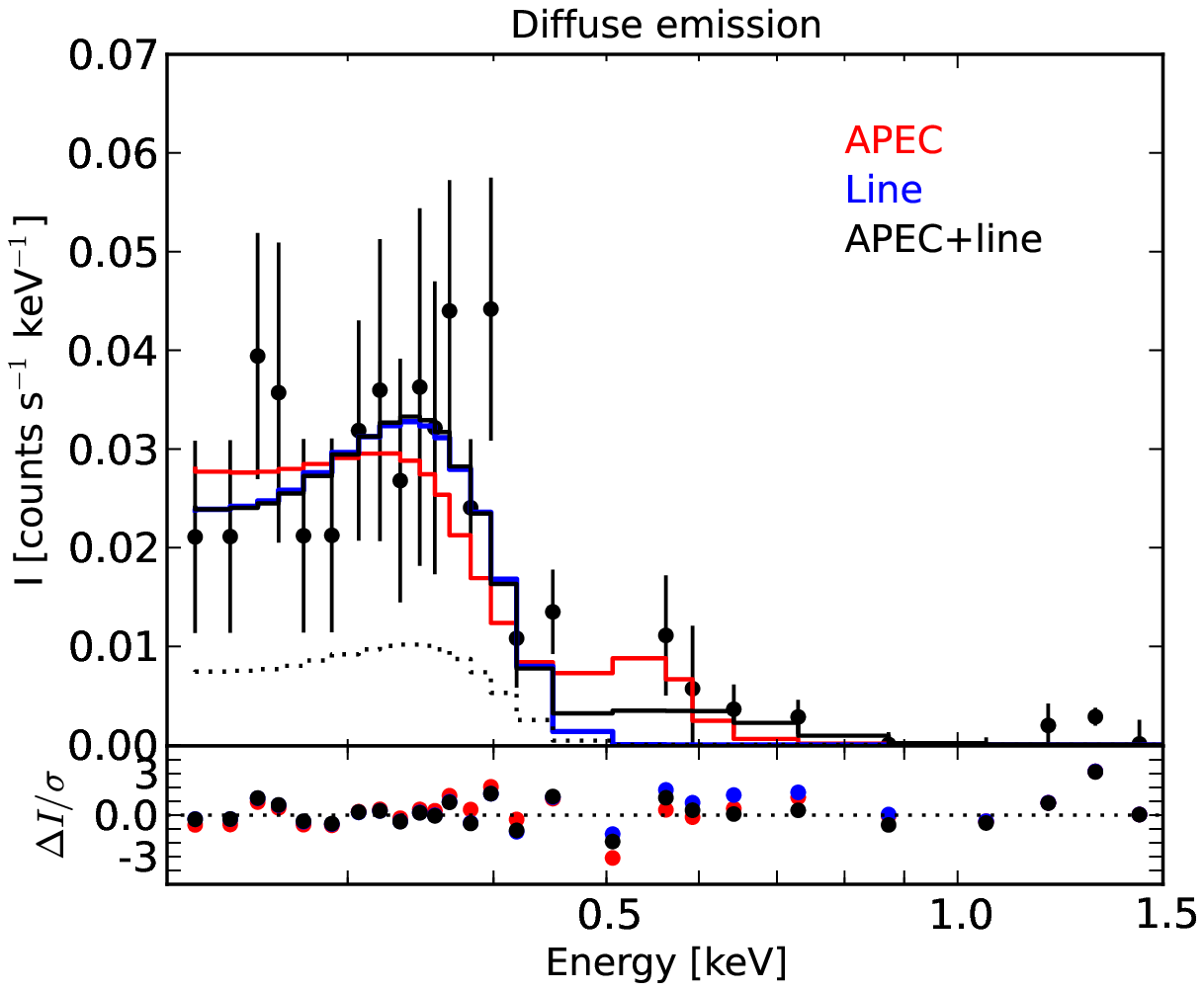}
\caption{
EPIC-pn background-subtracted spectra of the CSPN (left) and diffuse 
emission (right) of A78. 
Different line colors represent the different fits described in Table~2. 
The dotted histogram in the right panel illustrates the contribution 
of the CSPN to the diffuse emission.  
}
\end{center}
\label{fig:A78_radia}
\end{figure*}

\subsection{Spectral analysis}

The spectral analysis of the X-ray emission from A78 was performed
using XSPEC v.12.7.0 \citep{Arnaud1996}. The spectral fits include one
or two of the following components: (1) an emission line at $\sim$0.37
keV, and (2) an \emph{apec} optically thin plasma emission model with
abundances as those listed in Table~\ref{tab:CSPN}. The estimates of
the luminosity and electron density assume a distance of 1.4~kpc (see
\S2).

The emission model needs to be absorbed by the material along the line 
of sight whose origin can be interstellar, but also circumstellar.  
\citet{MP88} measured a nebular extinction towards the dusty, 
inner filaments of $E_{B-V}=0.15$ mag, which is in agreement 
with the UV absorption towards the CSPN derived in \S~2, but 
these measurements do not help disentangle the relative 
contributions of interstellar and circumstellar absorptions.  
To determine these contributions, we have used long-slit 
intermediate-dispersion optical spectroscopic observations 
(Fang et al., in preparation) of the outer ellipsoidal 
shell to derive the interstellar extinction towards the 
outer nebular regions.  
Our measurements indicate a negligible extinction, $E_{B-V}=0.014$ mag, 
thus proving that most of the absorption towards the central regions of 
A78 has a circumstellar origin, i.e., as in A30.  
The absorption would be produced by the dust in the central regions of A78.  
Since the IR emission of the dust \citep{Kimeswenger1998,Phillips2007} shares 
the same spatial distribution of the H-poor knots, the chemical composition 
of the dust is expected to be similar to that of the H-poor knots.  
Given the large metal-to-hydrogen ratio of this absorbing material, a
relatively small hydrogen column density,
$N_{\mathrm{H}}\approx$2$\times$10$^{16}$ cm$^{-2}$, will contain
similar amounts of carbon and oxygen as the interstellar hydrogen 
column density required to produce the observed absorption.

As a first inspection, the spectra of the three EPIC cameras were fit
simultaneously using an \textit{apec} component and an emission line
at 0.37~keV. This gives a good quality fit ($\chi^{2}$=72.6/77=1.06)
with a plasma temperature of $kT$=0.072 keV and $\Delta E$=26
eV\footnote{It is worth noting that, in the cases in which an emission
  line is used to model the X-ray emission, its line width is a few
  tens of eV, which means that the line is not resolved by
  EPIC-pn.}. The model is compared to the observed spectra in the
0.2--1.5~keV energy range in Figure~9-{\it top}, and the plasma
temperature ($kT$), fluxes ($f$), emission line central energy ($E$),
and normalization factors ($A$) of the best-fit models are listed in
Table~2.
The absorbed flux and intrinsic luminosity in the 0.2--2.0~keV energy 
range of this model are $f_{\mathrm{TOT}}$=(2.2$\pm$0.8)$\times$10$^{-14}$ 
erg~cm$^{-2}$~s$^{-1}$ and $L_{\mathrm{TOT,X}}$=(2.1$\pm$0.6)$\times$10$^{31}$ 
erg~s$^{-1}$.

Another emission model consisting of an \textit{apec} component and
two emission lines at 0.37~keV and 0.58~keV was tried in an attempt to
reproduce the emission excess at $\sim$0.58 keV, but it did not
improve statistically the results of the previous fit.  A simpler
model, involving just one \emph{apec} component, resulted in a poorer
fit, $\chi^{2}$/dof\footnote{
The degrees of freedom, dof, is equal to the number of spectral channels 
which are used into the fit minus the number of free parameters in the 
adopted model.  
}=70.26/45=1.56, for a similarly low plasma temperature
of $kT$=0.086 keV.

\begin{deluxetable*}{lcccccccr}
\tablewidth{0pt}
\tablecaption{Best-fit models for the X-ray emission in A78\tablenotemark{*}}
\tablehead{
\multicolumn{1}{l}{Region} &
\multicolumn{1}{c}{Model} &
\multicolumn{1}{c}{$kT$}  &
\multicolumn{1}{c}{$A_{1}$\tablenotemark{**}}  &
\multicolumn{1}{c}{$E$}  &
\multicolumn{1}{c}{$\Delta E$}  &
\multicolumn{1}{c}{$A_{2}$\tablenotemark{**}}  &
\multicolumn{1}{c}{$f_{\mathrm{X}}$}  &
\multicolumn{1}{c}{$\chi^{2}$/dof}  \\
\multicolumn{1}{l}{} &
\multicolumn{1}{c}{} &
\multicolumn{1}{c}{(keV)}  &
\multicolumn{1}{c}{(cm$^{-5}$)}  &
\multicolumn{1}{c}{(keV)}  &
\multicolumn{1}{c}{(keV)}  &
\multicolumn{1}{c}{(cm$^{-5}$)}  &
\multicolumn{1}{c}{(erg~cm$^{-2}$~s$^{-1}$)} &
\multicolumn{1}{c}{} 
 }
\startdata
A78    & line$+$apec & 0.072$^{+0.017}_{-0.013}$ & 1.60$\times$10$^{-9}$ & 0.37 & 2.6$\times$10$^{-2}$       &  6.7$\times$10$^{-5}$   & 2.20$\times$10$^{-14}$ & 72.58/72=1.06 \\
\hline 
CSPN   & apec & 0.071$^{+0.003}_{-0.003}$ & 1.60$\times$10$^{-9}$ & \dots & \dots                 &  \dots                 & 1.1$\times$10$^{-14}$     & 35.18/17=2.07 \\
       & line & \dots                   & \dots                 & 0.363$^{+0.017}_{-0.018}$ & 3.0$\times$10$^{-2}$  &  7.3$\times$10$^{-5}$  & 1.1$\times$10$^{-14}$ & 20.78/16=1.30 \\
       & line$+$apec & 1--64                   & 4.20$\times$10$^{-10}$ & 0.368$^{+0.014}_{-0.010}$ & 2.1$\times$10$^{-2}$  &  6.6$\times$10$^{-5}$  & 1.3$\times$10$^{-14}$ & 9.03/14=0.65 \\
\hline
Diffuse emission & apec & 0.086$^{+0.010}_{-0.012}$ & 7.20$\times$10$^{-10}$ & \dots                  & \dots                &  \dots               & 9.7$\times$10$^{-15}$ & 33.80/24=1.40 \\
                 & line & \dots                   & \dots                 & 0.362$^{+0.025}_{-0.023}$ & 3.6$\times$10$^{-2}$  &  5.8$\times$10$^{-5}$ & 8.8$\times$10$^{-15}$ & 31.75/23=1.38 \\
                 & line$+$apec & 1--40                   & 5.95$\times$10$^{-11}$ & 0.37 & 1.3$\times$10$^{-4}$  &  4.6$\times$10$^{-5}$ & 1.0$\times$10$^{-14}$ & 29.09/22=1.32 \\
\hline
Diffuse\\$+$CSPN spillover\tablenotemark{***} & diffuse$+$line &0.088$^{+0.019}_{-0.017}$ & 5.20$\times$10$^{-10}$ & 0.37 & 2.3$\times$10$^{-2}$  &  1.4$\times$10$^{-5}$ & 9.6$\times$10$^{-15}$ & 31.38/23=1.36 
\enddata
\label{tab:bestfit} 
\tablenotetext{*}{All models were computed assuming an absorbing column density of $N_\mathrm{H}=2\times10^{16}$~cm$^{-2}$.}
\tablenotetext{**}{The normalization parameter $A$ is defined as
  $A$=1$\times$10$^{-14} \int n_{\mathrm{e}}n_{\mathrm{H}} dV/4 \pi
  d^{2}$, where $d$ is the distance, $n_{\mathrm{e}}$ is the electron
  number density, and $V$ the volume, all in cgs units.}
\tablenotetext{***}{The line component is for the spillover CSPN emission.}
\end{deluxetable*}

We next proceeded to perform spectral fits of the X-ray emission from the 
CSPN and from the diffuse component separately.  
The different emission models used to describe these components are 
described in the next sections.  
We want to emphasize that the spectrum of the diffuse emission is 
contaminated by emission from the CSPN. 
This is not the case for the spectrum of the CSPN, which corresponds mostly 
to emission from the CSPN.

\subsubsection{X-ray emission from the CSPN}

The X-ray emission from the central star in A78 was first modeled by
an \textit{apec} plasma model. 
This resulted in a fit with reduced $\chi^{2}$ greater than two and plasma 
emission model of $kT$=0.071 keV. 
The absorbed flux in the 0.2--2.0~keV energy range is
$f_{\mathrm{X}}$=(1.08$\pm$0.35)$\times$10$^{-14}$ erg~cm$^{-2}$~s$^{-1}$.

A second fit was performed using only the contribution of an emission
line around $\sim$0.37~keV.  
This model results in an improved fit with $\chi^{2}$/dof=1.30. 
The corresponding absorbed flux is
(1.12$\pm$0.20)$\times$10$^{-14}$~erg~cm$^{-2}$~s$^{-1}$, very similar 
to the flux level derived for the single \emph{apec} model described 
above. 
The intrinsic X-ray luminosity of this model is 
(9.9$\pm$2.4)$\times$10$^{30}$ erg~s$^{-1}$.

A third model including the combination of an \textit{apec} model and an 
emission line at 0.37~keV model was also attempted, but this model does 
not improve the previous fits ($\chi^{2}$/dof=0.65). 
Indeed, XSPEC cannot constrain the temperature of the \textit{apec} 
component.

The three resultant models are compared with the background-subtracted
spectrum in the 0.2-1.5~keV energy range in Figure~10-left.  
The parameters of the best-fit models are summarized in Table~2.  

\subsubsection{The diffuse and extended X-ray emission}

Similarly, the diffuse X-ray emission in A78 was first fit using an 
\emph{apec} optically thin plasma model.  
The fit has a reduced $\chi^{2}$ of 1.40, for a plasma temperature of 
$kT$=0.086 keV and an absorbed flux of 
(9.70$\pm$0.25)$\times$10$^{-15}$~erg~cm$^{-2}$~s$^{-1}$. 

The second fit adopted a single emission line at $\sim$0.37~keV. This
model resulted in a similar quality fit, with $\chi^{2}$/dof of 1.38,
for an emission line with energy 0.362~keV and $\Delta E$=36~eV. The
corresponding absorbed flux is (8.8$\pm$0.5)$\times$10$^{-15}$
erg~cm$^{-2}$~s$^{-1}$.

The third model used a combination of an \emph{apec} plasma model and
an emission line at 0.37~keV. This model gives a reduced $\chi^{2}$
of 1.32, but the plasma temperature is basically unconstrained and the
line width implies it is unresolved. Still, the resultant absorbed
flux, (1.0$\pm$0.6)$\times$10$^{-14}$ erg~cm$^{-2}$~s$^{-1}$, is
consistent with those implied by the other two models.

One final spectral fit was attempted taking into account that the CSPN 
contributes significantly to the emission registered in this region.  
This contribution was estimated to be $\approx$1/4 of the emission 
detected in the region defined for the CSPN.  
We then scaled the CSPN line emission model listed in Table~2 to 
this flux level and added it as an additional spectral component 
to the model of the diffuse emission, which consisted of an 
\emph{apec} model.  
The best-fit parameters of this model are listed in the last row of Table~2.  
The quality of the fit is similar to the previous ones, but the temperature 
of the plasma is constrained to $kT$=0.088 keV.  
Its corresponding flux is
(9.6$\pm$1.6)$\times$10$^{-15}$~erg~cm$^{-2}$~s$^{-1}$, but only
(7.4$\pm$1.7)$\times$10$^{-15}$~erg~cm$^{-2}$~s$^{-1}$ corresponds to
the diffuse X-ray emission.
The intrinsic X-ray luminosity of the net diffuse emission is 
(6.8$\pm$1.7)$\times$10$^{30}$ erg~s$^{-1}$.

The best-fit models are compared to the background-subtracted spectrum
from the diffuse emission in the 0.2--1.5~keV in Figure~10-right, and the 
parameters listed in Table~2.\\
\\

\section{Discussion}
\label{sec:discussion}

The present \textit{XMM-Newton} observations have discovered very soft
X-ray emission in the born-again PN A78. These X-ray observations
reveal the existence of a point-like source associated with the CSPN
and a source of extended X-ray emission within A78. The spatial
distribution of the diffuse X-ray emission does not fill the
elliptical outer shell; instead, this emission can be associated with
the H-poor knots and is enclosed by the filamentary cavity detected in
[O~{\sc iii}] narrow band images.

The best-fit model for the CSPN of A78 seems to be that including only
a line emission at $\sim$0.37 keV; models including a plasma component
cannot produce an acceptable fit to the data. The estimated total flux
for the CSPN, accounting for the fraction of the PSF not included in
the source aperture of radius 12\arcsec, is
$f_{\mathrm{X,CSPN}}$=(1.32$\pm$0.24)$\times$10$^{-14}$ erg~cm$^{-2}$~s$^{-1}$
which corresponds to an intrinsic X-ray luminosity of
$L_{\mathrm{X,CSPN}}$=(1.2$\pm$0.3)$\times$10$^{31}$ erg~s$^{-1}$.

The X-ray emission from the diffuse component in A78 is better
explained by a thermal plasma with temperature of $kT$=0.088 keV 
($T\approx$1.0$\times$10$^{6}$ K).  
Its corresponding total flux, after subtracting the contribution from 
the CSPN and adding that of regions with aperture radius $<$17\arcsec, is
$f_{\mathrm{X,DIFF}}$=(1.0$\pm$2.2)$\times$10$^{-14}$ erg~cm$^{-2}$~s$^{-1}$,
which corresponds to an intrinsic luminosity of
$L_{\mathrm{X,DIFF}}$=(9.2$\pm$2.3)$\times$10$^{30}$ erg~s$^{-1}$.
The normalization parameter for the \textit{apec} component
($A_{1}$=5.20$\times$10$^{-10}$ cm$^{-5}$) has been used to estimate
an electron density of the diffuse X-ray-emitting gas for a distance
of 1.4~kpc as
$n_{\mathrm{e}}$=0.002 $(\epsilon/0.1)^{1/2}$ cm$^{-3}$,
with $\epsilon$ as the gas-filling factor.

\subsection{Comparison with A30}

The four {\it bona fide} born-again PNe, A30, A58, A78, and Sakurai's
object, represent different stage of the same evolutionary path. 
A30 and A78 are very similar in many ways. The morphology and spectral
similarities between A30 and A78 and their central stars are
remarkable. Their optical narrow band images show similar
limb-brightened outer nebulae which correspond to the expected shell
in the canonical formation of a PN \citep{Kwok1978,Balick1987}, with
estimated dynamical ages, $\tau_\mathrm{dyn}$\footnote{The dynamical
  age can be estimated as $\tau_\mathrm{dyn}=R/v_\mathrm{exp}$, where
  $R$ and $v_\mathrm{exp}$ are the radius and velocity of the outer
  optical shell, respectively.}, of 12,500 and 10,700~yr for A30 and
A78, respectively.  The processed H-poor material is thought to have
been ejected around a thousand years ago
\citep[e.g.,][]{Guerrero2012,Fang2014}, which means that the stellar
wind velocity must have increased very rapidly within this time-lapse
in both cases.  The interaction of this stellar wind with the material
ejected in the born-again event is responsible for the shaping of the
cloverleaf and eye-shaped H-poor clumpy distributions in A30 and A78,
respectively \citep[see][]{Fang2014}.

The X-ray properties of A30 and A78 are very much alike. The diffuse X-ray
emission can be modeled by an optically thin plasma model with
similarly low temperatures for both PNe, besides the different
relative importance of an emission line at $\sim$0.58 keV attributable
to O~{\sc vii} in their spectra.  The origin of this hot plasma may be
due to pockets of shocked and thermalized stellar wind, as the current
fast wind from the CSPNe ($V_{\infty}\lesssim$4000~km~s$^{-1}$)
interacts with the processed material from the born-again ejection.
The plasma temperature from an adiabatic shocked stellar wind can be
determined by $kT=3\mu m_{\mathrm{H}} V_{\infty}^{2}/16$, where $\mu$
is the mean particle mass \citep{Dyson1997}.  Therefore, for the
stellar winds of A30 and A78, the temperature expected from the
shocked material would be $T\sim$5$\times$10$^8$~K, in sharp contrast
with the observed temperatures.  This discrepancy is found in all PNe
in which diffuse X-ray emission is detected \citep[see][and references
  therein]{Ruiz2013} and it is always argued that thermal conduction
is able to reduce the temperature of the shocked stellar wind and to
increase its density to observable values \citep{Soker1994}. Even
though one-dimension radiative-hydrodynamic models on the formation of
hot bubbles in PNe as those presented by \citet{Steffen2008} and
\citet{Steffen2012} are able to explain this discrepancy, they are not
tailored to the specific evolution of a star that experiences a VLTP
and creates a born-again PN. The fact that the diffuse X-ray emission
of A30 and A78 is confined within filamentary and clumpy H-poor shells
is a clear indication that a variety of physical processes are taking
place to reduce its temperature. These processes may include
mass-loading and photoevaporation from the H-poor clumps
\citep{Meaburn2003}, which mix the material with the thermalized
shocked wind.  The realization of numerical simulations on the
formation of born-again PNe including these complex interactions and
accounting for their singular abundances is most needed to understand
the puzzling X-ray emission in these objects (Toal\'{a} \& Arthur in
prep.).

The X-ray emission from the point sources at the CSPNe of A30 and A78
is dominated by the C\,{\sc vi} line.  As discussed by
\citet{Guerrero2012}, the origin of the X-ray emission associated with
the CSPNe of born-again PNe is inconclusive. Several mechanism are
capable of producing this X-ray emission (e.g., charge transfer
reactions from highly ionized species of carbon, oxygen, and
nitrogen), but the present observations cannot provide a definite
answer.  

It is worth mentioning here that, given the large opacity of this
material \citep[see figure~11 in][]{Guerrero2012}, the C~{\sc vi}
0.37~keV emitting-region must be located at least a few stellar radii
above the surface of the CSPN, which is consistent with the
carbon-rich wind implied by our optical and UV spectral modeling.

\subsection{On the origin of the hydrogen-poor material}

The origin of the newly processed hydrogen-poor material inside the
old PN is still a matter of debate \citep{Lau2011}.  
The Ne abundance of the ejecta can hold important clues.  
The Ne abundances of A30 and A58 \citep{Wesson2003,Wesson2008} 
seem to point to a nova eruption on an O-Ne-Mg WD \citep{Lau2011}. 
This hypothesis was invoked by \citet{Maness2003} to interpret the 
high Ne abundances in the X-ray-emitting gas in BD+30$^\circ$3639.

That is an interesting possibility worth to be explored in A78.  
First, we have to note that there is no significant contribution of 
Ne lines ($\sim$0.9~keV) to the observed spectrum of A78.
This does not necessarily imply a low Ne abundance, because the 
temperature of the hot plasma in A78 is low to produce bright 
Ne~{\sc ix} and Ne~{\sc x} emission lines.  
To test this, we have compared the observed X-ray spectrum with models 
with Ne abundance enhanced by 5, 10, and 20 times the value reported in 
\S2, where the mid values in this range implies Ne abundances by mass 
similarly high to those reported in A30 and A58.  
These changes in the abundances do not produce noticeable effects 
in the synthetic spectrum, revealing a fundamental insensitivity 
of the X-ray spectrum of A78 to the Ne abundance due to its low 
plasma temperature.

\section{Summary and conclusions}

We report the \emph{XMM-Newton} discovery of X-ray emission in A78, 
making it the second born-again PN detected in X-rays, besides A30 
\citet{Guerrero2012}.  
The X-ray data of A78 have been analyzed in conjunction with narrow-band 
optical images of the nebula to determine the spatial distribution of the 
hot gas.  
Multiwavelength spectral observations of its CSPN have also been analyzed 
using a NLTE code for expanding atmospheres to assess its stellar parameters 
and wind properties.  
In particular we find:

\begin{itemize}

\item The spatial distribution and spectral properties of the X-ray
  emission detected towards A78, and the chemical abundances and
  stellar wind parameters of its CSPN are very similar to those
  reported for A30 by \citet{Guerrero2012}.

\item The X-ray emission from A78 consists of a point-like source and
  diffuse emission.  The point-like source is coincident with the
  position of the CSPN.  The distribution of diffuse X-ray emission
  does not fill the outer nebular shell, instead it traces the H-poor
  clumps and eye-shaped cavity detected in [O~{\sc iii}] narrow band
  images.  An apparent maximum in the diffuse X-ray emission is
  detected at the location of one H-poor clump toward the southwest.

\item The X-ray emission from A78 is very soft.  Most of the X-ray
  emission has energies lower than 0.5 keV, pointing at the C~{\sc vi}
  0.37 keV emission line.  There is evidence for a weaker emission
  line at $\sim$0.58 keV, which would correspond to the O~{\sc vii}
  triplet.

\item The analysis of the optical and UV spectra of the CSPN of A78
  helped us to constrain its abundances and stellar wind parameters.
  These have been used for the analysis of the X-ray spectra of A78.

\item The best-fit model for the diffuse X-ray emission resulted in a
  plasma temperature of $kT$=0.088 keV ($T\approx$1.0$\times$10$^6$ K)
  with an estimated absorbed flux of
  $f_{\mathrm{X,DIFF}}$=1.0$\times$10$^{-14}$ erg~cm$^{-2}$~s$^{-1}$.
  The estimated X-ray luminosity is
  $L_{\mathrm{X,DIFF}}$=9.2$\times$10$^{30}$ erg~s$^{-1}$.  A variety
  of processes may have played significant roles in lowering the
  plasma temperature (e.g., mixing, ablation and photoevaporation) of
  the diffuse X-ray emission.

\item The X-ray spectra of A78 cannot be used to constrain the Ne
  abundance of the hot plasma due to its low temperature.

\item The main X-ray spectral feature in the CSPN in A78 is the C~{\sc
    vi} emission line, as revealed by the EPIC and RGS spectra.  Its
  estimated flux in the 0.2--2.0~keV energy band is
  $f_{\mathrm{X,CSPN}}$=1.32$\times$10$^{-14}$ erg~cm$^{-2}$~s$^{-1}$
  which corresponds to a luminosity of
  $L_{\mathrm{X,CSPN}}$=1.2$\times$10$^{31}$ erg~s$^{-1}$.  The
  physical mechanism for the production of the emission associated
  with the CSPN of A78 is elusive, as it is also the case for the CSPN
  of A30.

\end{itemize}

\acknowledgements We would like to thank the referee, Orsola De Marco,
for her valuable comments and suggestions. We also thank K.\,Werner
for fruitful discussion. J.A.T. acknowledges support by the CSIC
JAE-Pre student grant 2011-00189. J.A.T., M.A.G., R.A.M.-L., and
X.F. are supported by the Spanish MICINN (Ministerio de Ciencia e
Innovaci\'on) grant AYA 2011-29754-C03-02. We acknowledge support from
DLR grant 50 OR 1302.


\begin{thebibliography}{}

\bibitem[Althaus et al.(2005)]{Althaus2005} Althaus, L.~G., Serenelli,
  A.~M., Panei, J.~A., et al.\ 2005, \aap, 435, 631

\bibitem[Arnaud(1996)]{Arnaud1996} Arnaud, K.~A.\ 1996, Astronomical
  Data Analysis Software and Systems V, 101, 17

\bibitem[Asplund et al.(1999)]{Asplund1999}
Asplund, M., Lambert, D.L., Kipper, T., Pollacco, D., \& Shetrone, M.D.\ 
1999, \aap, 343, 507

\bibitem[Balick(1987)]{Balick1987} 
Balick, B.\ 
1987, \aj, 94, 671 

\bibitem[Borkowski et al.(1993)]{Borkowski1993} Borkowski, K.~J.,
  Harrington, J.~P., Tsvetanov, Z., \& Clegg, R.~E.~S.\ 1993, ApJL,
  415, L47

\bibitem[Cardelli et al.(1989)]{Cardelli1989} Cardelli, J.~A.,
  Clayton, G.~C., \& Mathis, J.~S.\ 1989, \apj, 345, 245

\bibitem[Chu et al.(1997)]{Chu1997} Chu, Y.-H., Chang, T.~H., \&
  Conway, G.~M.\ 1997, \apj, 482, 891

\bibitem[Chu \& Ho(1995)]{Chu1995} Chu, Y.-H., \& Ho, C.-H.\ 1995,
  \apjl, 448, L127

\bibitem[Clayton et al.(2013)]{Clayton2013} 
Clayton, G.~C., Bond, H.~E., Long, L.~A., et al.\ 2013, 
\apj, 771, 130 

\bibitem[Dyson \& Williams(1997)]{Dyson1997} Dyson, J.~E., \&
  Williams, D.~A.\ 1997, The physics of the interstellar medium.~
  Edition: 2nd ed.~Publisher: Bristol: Institute of Physics
  Publishing, 1997.~Edited by J.~E.~Dyson and D.~A.~Williams.~Series:
  The graduate series in astronomy.~ISBN: 0750303069

\bibitem[Evans et al.(2006)]{Evans2006} Evans, A., Tyne, V.~H., van
  Loon, J.~T., et al.\ 2006, \mnras, 373, L75


\bibitem[Fang et al.(2014)]{Fang2014} Fang, X., Guerrero, M.\,A.,
  Marquez-Lugo, R.\,A., et al., 2014, accepted to ApJ, arXiv:1410.3872v1

\bibitem[Freeman et al.(2014)]{Freeman2014} Freeman, M., Montez, R.,
  Jr., Kastner, J.~H., et al.\ 2014, \apj, 794, 99

\bibitem[Frew(2008)]{Frew2008} Frew, D.~J.\ 2008, Ph.D.~Thesis

\bibitem[Gr{\"a}fener et al.(2002)]{Grafener2002} Gr{\"a}fener,
  G., Koesterke, L., \& Hamann, W.-R.\ 2002, \aap, 387, 244

\bibitem[Guerrero \& De Marco(2013)]{Guerrero2013} Guerrero, M.~A., \&
  De Marco, O.\ 2013, \aap, 553, A126


\bibitem[Guerrero et al.(2012)]{Guerrero2012} Guerrero, M.~A., Ruiz,
  N., Hamann, W.-R., et al.\ 2012, \apj, 755, 129

\bibitem[Hamann \& Gr{\"a}fener(2004)]{Hamann2004} Hamann, W.-R., \&
  Gr{\"a}fener, G.\ 2004, \aap, 427, 697

\bibitem[Harrington et al.(1995)]{Harrington1995} Harrington, J.~P.,
  Borkowski, K.~J., \& Tsvetanov, Z.\ 1995, \apj, 439, 264

\bibitem[Herald et al.(2005)]{Herald2005} Herald, J.~E., Bianchi, L.,
  \& Hillier, D.~J.\ 2005, \apj, 627, 424

\bibitem[Herwig et al.(1999)]{Herwig1999} Herwig, F., Bl{\"o}cker, T.,
  Langer, N., \& Driebe, T.\ 1999, \aap, 349, L5

\bibitem[Hinkle \& Joyce(2014)]{Hinkle2014} Hinkle, K.~H., \& Joyce,
  R.~R.\ 2014, \apj, 785, 146

\bibitem[Hinkle et al.(2008)]{Hinkle2008} Hinkle, K.~H., Lebzelter,
  T., Joyce, R.~R., et al.\ 2008, A\&A, 479, 817

\bibitem[Jacoby(1979)]{Jacoby1979} Jacoby, G.~H.\ 1979, \pasp, 91, 754

\bibitem[Jeffery(1995)]{Jeffery1995} Jeffery, C.~S.\ 1995, \aap, 299,
  135

\bibitem[Kastner et al.(2012)]{Kastner2012} Kastner, J.~H., Montez,
  R., Jr., Balick, B., et al.\ 2012, \aj, 144, 58

\bibitem[Kimeswenger et al.(1998)]{Kimeswenger1998} Kimeswenger, S.,
  Kerber, F., \& Weinberger, R.\ 1998, \mnras, 296, 614

\bibitem[Koesterke \& Werner(1998)]{Koesterke1998} Koesterke, L., \&
  Werner, K.\ 1998, \apjl, 500, L55

\bibitem[Kwok et al.(1978)]{Kwok1978} Kwok, S., Purton, C.~R., \&
  Fitzgerald, P.~M.\ 1978, \apjl, 219, L125

\bibitem[Lau et al.(2011)]{Lau2011} Lau, H.~H.~B., De Marco, O., \&
  Liu, X.-W.\ 2011, \mnras, 410, 1870

\bibitem[Lawlor \& MacDonald(2006)]{Lawlor2006} Lawlor, T.~M., \&
  MacDonald, J.\ 2006, \mnras, 371, 263

\bibitem[Leuenhagen et al.(1993)]{Leue1993} Leuenhagen, U., Koesterke,
  L., \& Hamann, W.-R.\ 1993, Acta Astron., 43, 329

\bibitem[Manchado et al.(1988)]{MP88} Manchado, A., Mampaso, A., \&
  Pottasch, S.~R.\ 1988, \aap, 191, 128

\bibitem[Maness et al.(2003)]{Maness2003} Maness, H.~L., Vrtilek,
  S.~D., Kastner, J.~H., \& Soker, N.\ 2003, \apj, 589, 439

\bibitem[Meaburn \& Redman(2003)]{Meaburn2003} Meaburn, J., \& Redman,
  M.~P.\ 2003, Revista Mexicana de Astronomia y Astrofisica Conference
  Series, 15, 1

\bibitem[Meaburn et al.(1998)]{Meaburn1998} Meaburn, J., Lopez, J.~A.,
  Bryce, M., \& Redman, M.~P.\ 1998, \aap, 334, 670

\bibitem[Meaburn \& Lopez(1996)]{Meaburn1996} Meaburn, J., \& Lopez,
  J.~A.\ 1996, \apjl, 472, L45

\bibitem[Miller Bertolami \& Althaus(2007)]{MillerBertolami2007}
  Miller Bertolami, M.~M., \& Althaus, L.~G.\ 2007, \mnras, 380, 763

\bibitem[Miller Bertolami \& Althaus(2006)]{Millerbertolami2006a}
  Miller Bertolami, M.~M., \& Althaus, L.~G.\ 2006, \aap, 454, 845

\bibitem[Miller Bertolami et al.(2006)]{Millerbertolami2006b} Miller
  Bertolami, M.~M., Althaus, L.~G., Serenelli, A.~M., \& Panei,
  J.~A.\ 2006, \aap, 449, 313


\bibitem[Phillips \& Ramos-Larios(2007)]{Phillips2007} Phillips,
  J.~P., \& Ramos-Larios, G.\ 2007, \aj, 133, 347

\bibitem[Pottasch(1984)]{Pottasch1984} Pottasch, S.~R.\ 1984,
  Astrophysics and Space Science Library, Vol. 107, Planetary Nebulae
  - A study of late stages of stellar evolution


\bibitem[Ruiz et al.(2013)]{Ruiz2013} Ruiz, N., Chu, Y.-H., Gruendl,
  R.~A., et al.\ 2013, \apj, 767, 35

\bibitem[Sch{\"o}nberner et al.(2005)]{Schonberner2005}
  Sch{\"o}nberner, D., Jacob, R., Steffen, M., et al.\ 2005, \aap,
  431, 963

\bibitem[Steffen et al.(2012)]{Steffen2012} Steffen, M., Sandin, C.,
  Jacob, R., \& Sch{\"o}nberner, D.\ 2012, IAU Symposium, 283, 215

\bibitem[Steffen et al.(2008)]{Steffen2008} Steffen, M.,
  Sch{\"o}nberner, D., \& Warmuth, A.\ 2008, \aap, 489, 173

\bibitem[Soker(1994)]{Soker1994} Soker, N.\ 1994, \aj, 107, 276

\bibitem[Tarafdar \& Apparao(1988)]{Tarafdar1988} Tarafdar, S.~P., \&
  Apparao, K.~M.~V.\ 1988, ApJ, 327, 342


\bibitem[Werner \& Koesterke(1992)]{Werner1992} Werner, K., \&
  Koesterke, L.\ 1992, The Atmospheres of Early-Type Stars, 401, 288

\bibitem[Werner et al.(2005)]{Werner2005} Werner, K., Rauch, T., \&
  Kruk, J.~W.\ 2005, \aap, 433, 641

\bibitem[Werner et al.(2011)]{Werner2011} Werner, K., Rauch, T., Kruk,
  J.~W., \& Kurucz, R.~L.\ 2011, \aap, 531, A146

\bibitem[Wesson et al.(2008)]{Wesson2008} Wesson, R., Barlow, M.~J.,
  Liu, X.-W., et al.\ 2008, \mnras, 383, 1639

\bibitem[Wesson et al.(2003)]{Wesson2003} Wesson, R., Liu, X.-W., \&
  Barlow, M.~J.\ 2003, \mnras, 340, 253

\bibitem[Zsarg{\'o} et al.(2008)]{Zsargo2008} Zsarg{\'o}, J., Hillier,
  D.~J., Bouret, J.-C., et al.\ 2008, \apjl, 685, L149

\end{thebibliography}
\end{document}